\documentclass[letterpaper,10pt, conference]{ieeeconf}

\IEEEoverridecommandlockouts          

\usepackage{graphicx}
\usepackage{subfigure}
\usepackage{caption}
\usepackage{amsmath}
\usepackage{amssymb}
\usepackage{accents}
\usepackage{statex}
\usepackage{cases}
\usepackage{mathrsfs}
\usepackage{tikz,pgfplots}
\pgfplotsset{compat=newest}
\usepackage{xcolor}
\usepackage{caption}
\usepackage{setspace}
\usepackage{booktabs}
\usepackage{array}
\usepackage{cite}
\let\labelindent\relax
\usepackage{enumitem}
\usepackage{algorithm}
\usepackage{algorithmic} 
\usepackage{multirow}
\usetikzlibrary{patterns}
\usetikzlibrary{graphs}
\usetikzlibrary{graphs.standard}
\usetikzlibrary{matrix,arrows,fit,backgrounds,mindmap,plotmarks,decorations.pathreplacing}
\newcolumntype{C}[1]{>{\centering\arraybackslash}p{#1}}
\usepackage{eqparbox}

\usetikzlibrary{automata}

\newtheorem{lemma}{Lemma}

\newlength\figureheight
\newlength\figurewidth

\DeclareMathOperator*{\argmin}{argmin}

\DeclareMathOperator*{\mD}{\mathcal{D}}

\title{\bf MPC of Uncertain Nonlinear Systems with Meta-Learning for Fast Adaptation of Neural Predictive Models}

\author{Jiaqi Yan, Ankush Chakrabarty, Alisa Rupenyan, and John Lygeros
	\thanks{
		J. Yan and J. Lygeros are with Automatic Control Laboratory, ETH Zurich, Switzerland (\{jiayan,jlygeros\}@ethz.ch).
		A. Chakrabarty is with Mitsubishi Electric Research Laboratories, Cambridge, MA 02139, USA (achakrabarty@ieee.org).
		A. Rupenyan is with ZHAW Centre for Artificial Intelligence, Zurich University of Applied Sciences, Switzerland  (alisa.rupenyan@zhaw.ch).		
		This work is supported by the Swiss National Science Foundation supported this work through NCCR Automation under Grant agreement 51NF40\_180545.
	}
}

\begin{document}
	\maketitle
	
	\begin{abstract}
	In this paper, we consider the problem of reference tracking in uncertain nonlinear systems. A neural State-Space Model (NSSM) is used to approximate the nonlinear system, where a deep encoder network learns the nonlinearity from data, and a state-space component captures the temporal relationship. This transforms the nonlinear system into a linear system in a latent space, enabling the application of model predictive control (MPC) to determine effective control actions. Our objective is to design the optimal controller using limited data from the \textit{target system} (the system of interest). To this end, we employ an implicit model-agnostic meta-learning (iMAML) framework that leverages information from \textit{source systems} (systems that share similarities with the target system) to expedite training in the target system and enhance its control performance. The framework consists of two phases: the (offine) meta-training phase learns an aggregated NSSM using data from source systems, and the (online) meta-inference phase quickly adapts this aggregated model to the target system using only a few data points and few online training iterations, based on local loss function gradients. The iMAML algorithm exploits the implicit function theorem to exactly compute the gradient during training, without relying on the entire optimization path. By focusing solely on the optimal solution, rather than the path, we can meta-train with less storage complexity and fewer approximations than other contemporary meta-learning algorithms. 
 We demonstrate through numerical examples that our proposed method can yield accurate predictive models by adaptation, resulting in a downstream MPC that outperforms several baselines.
	\end{abstract}

	\section{Introduction}

 Optimal control for unknown nonlinear systems has been a long-standing challenge in various applications such as robotics, manufacturing systems, and so on. Different learning-based approaches have been explored, such as reinforcement learning, adaptive dynamic programming, and stochastic optimization-based control \cite{dierks2012online,modares2017optimal,pepyne2000optimal}. 
Recently, the neural state-space model (NSSM) has gained popularity for tackling nonlinear control problems \cite{chakrabarty2023meta,legaard2023constructing}. This model enhances traditional state-space models by employing neural networks to capture nonlinearity of the system. 
The architecture is grounded in Koopman operator theory, demonstrating that any nonlinear system can be lifted to an infinite-dimensional linear system in a latent space \cite{koopman1932dynamical}. For the sake of tractable computations, the latent state of NSSMs is often chosen to be of finite size.

Although NSSMs show good performance in approximating nonlinear systems, their training usually requires a large dataset. In situations where data is scarce, constructing an accurate model becomes challenging. This difficulty often arises when collecting data from the system of interest, referred to as the \textit{target system}, is arduous or expensive.
In response to this challenge, recent efforts, while not specifically within the domain of NSSMs, aim to leverage data collected from \textit{source systems} that share similarities with the target system \cite{xin2022identifying,li2023data}. As an example, the source system may be a numerical model or a digital twin of the target system, or a system with modified physical parameters that is easily accessible, allowing for the collection of large amounts of data. We aim to use this shared dataset to pre-train a model, then fine-tune it on the target system with limited data.

This paper extends the approach of leveraging knowledge acquired from similar systems to expedite training of the target system in the context of NSSM. Different from \cite{xin2022identifying,li2023data} that use a transfer learning framework, we propose a method by using meta-learning. The distinction between transfer learning and meta-learning is elucidated in \cite{dumoulin2021comparing}. 

Meta-learning has recently gained widespread application in control and robotics, addressing challenges such as system identification, parameter estimation, and adaptive control \cite{toso2024meta,chakrabarty2023meta,zhan2022calibrating}. In most of these works, model-agnostic meta-learning (MAML) based solutions are developed, which involves solving a bilevel optimization problem in the meta-training stage. The inner loop of MAML involves calculating and propagating derivatives of the training loss function along the full optimization path (which is the same as the number of inner-loop updates/steps). The outer loop then accumulates information from these inner-loop updates and computes an outer-loop gradient direction. By repeating these two loops, the MAML learner is expected to asymptotically converge to a set of neural weights from which rapid adaptation is possible to any task within the distribution of training tasks. Choosing a large number of inner-loop steps, while useful to understand good outer-loop directions, incurs high memory complexity, and is not practical for a large number of inner-loop adaptation steps. Therefore, existing approaches typically propose approximations to obtain a more efficient solution. Unfortunately, these approximations are sometimes oversimplifications or only valid for specific activation functions (e.g., first-order MAML exploits ReLU properties), and thus, can harm quality. One particular approximation that utilizes the implicit function theorem is called implicit MAML (iMAML), that enables computing the asymptotic inner-loop gradient efficiently~\cite{rajeswaran2019meta}. By doing this, the meta-learner can update the outer-loop based on directions obtained by solving the inner-loop over a large number of iterations without requiring high memory, as the method is agnostic to the path taken in the inner-loop and only uses the final gradient direction.
As a result, better adaptation performance can be expected.

In this paper, we solve the optimal tracking problem in unknown nonlinear systems by leveraging the iMAML algorithm for system identification, and using this identified predictive model within a receding horizon framework. This is summarized herein:

(i) To efficiently utilize limited samples from the target system, we propose a meta-learning framework that customizes iMAML for adapting neural predictive models. The framework pre-trains an aggregated model from source systems, and then fine-tunes it with a small target dataset and few adaptation steps. (ii) We incorporate the adapted models into MPC, determining optimal inputs for the target system.  (iii) Numerical examples show that our method outperforms models trained solely by MAML or with only target system data, along with other baselines.

	
	\section{Problem Formulation}\label{sec:problem}
	We consider a family of parameterized discrete-time nonlinear systems of the form 
	\begin{equation}\label{eqn:sys}
		\begin{split}
			x_{t+1} &= f(x_t,u_t,\theta_f),\\
			y_t &= g(x_t,u_t,\theta_g),
		\end{split}
	\end{equation}
	where $x\in\mathbb{R}^n$ and $y\in\mathbb{R}^m$ are respectively system state and output. Moreover, $u\in \mathcal{C} \subseteq \mathbb{R}^p$ is the input with $\mathcal{C}$ being a compact, convex set. Both $f$ and $g$ are unknown nonlinear functions, and $\theta:=[\theta_f; \theta_g]\in\mathbb{R}^{w}$ denotes a vector of unknown parameters. 
	%
	 Despite the unknown dynamics, our objective is to design a controller to asymptotically track some reference signal $\bar{y}$ in the sense that
	\begin{equation}\label{eqn:objective}
	\lim_{t\to\infty} y_t = \bar{y}_{t}.
	\end{equation}
	

\subsection{Neural State-Space Model (NSSM)}
For the unknown system described by \eqref{eqn:sys}, we construct a neural state-space model (NSSM) that approximates the dynamics of the \textit{target system} (the system of interest) parameterized by $\theta =\theta^*$. The true value of $\theta^*$ is unknown, but it is drawn from a distribution 
$	
		\theta^* \sim \Theta,
$
where $\Theta$ is often informed by domain experience. Therefore, it is not impractical to assume we know $\Theta$, but it is not necessarily for our proposed method: we instead assume that $\Theta$ can be reliably sampled from even though its distributional form is unknown.
	
Let $U(\theta^*, T^*)$ and $Y(\theta^*, T^*)$ represent the input and output trajectories generated from the target system over a horizon $T^*$ and construct a target dataset 
\begin{equation}
	\mathcal{D}_{\text{target}} := \{U(\theta^*, T^*), Y(\theta^*, T^*)\}.
\end{equation}
Using this dataset, one can train an NSSM of the form
	\begin{subequations}\label{eqn:SSM}
		\begin{align}
			z_t &= f_{\text{enc}}(U_{t-H+1:t},Y_{t-H+1:t}), \label{eqn:z}\\
			z_{t+1} &= A_z z_{t}+ B_z u_{t+1},\label{eqn:z_next}\\ 
			\hat{y}_{t} &= C_z z_t.  \label{eqn:haty}
		\end{align}
	\end{subequations}
	where $z\in\mathbb{R}^{n_z}$ is the latent state learned by the encoder network $f_{\text{enc}}(\cdot)$ from historical input and output data:
	\begin{equation}
		\begin{split}
			U_{t-H+1:t}&:= \{u_{t-H+1},u_{u-H+2},\cdots,u_{t}\} \subseteq \mathcal{D}_{\text{target}},\\
			Y_{t-H+1:t}&:= \{y_{t-H+1},y_{t-H+2},\cdots,y_{t}\}\subseteq \mathcal{D}_{\text{target}}.
		\end{split}
	\end{equation}
	Here, the latent state dimension $n_z \in \mathbb{N}$ and window length $H\in \mathbb{N}$ are user-defined parameters; these are standard hyperparameters in classical system identification as well. Also, $\hat{y}_{t}\in\mathbb{R}^m$ is the predicted output, which we require to be close to the true output $y_{t}$ after NSSM training~\cite{chakrabarty2023meta}.

	Note that the NSSM \eqref{eqn:SSM} consists of two components: the deep latent encoder $f_{\text{enc}}$ in \eqref{eqn:z} captures the nonlinearity of the system, while the state-space component in \eqref{eqn:z_next}--\eqref{eqn:haty} models temporal dependencies in the system. 
Training \eqref{eqn:SSM} involves optimizing the weights of the encoder network $f_{\text{enc}}(\cdot)$, and the elements of the linear decoders $A_z, B_z, C_z$ and $D_z$; let $\omega$ be the set of these parameters. The prediction performance of \eqref{eqn:SSM} is evaluated as follows. 
	\begin{enumerate}
		\item We first construct a dataset $\mathcal{D}$ from $\mathcal{D}_{\text{target}}$ by collecting trajectories of length $H+T$:
		\begin{equation}\notag
		\begin{split}
		U_{t-H+1:t+T}&:= \{u_{t-H+1},u_{u-H+2},\cdots,u_{t+T}\},\\
		Y_{t-H+1:t+T}&:= \{y_{t-H+1},y_{t-H+2},\cdots,y_{t+T}\}.
		\end{split}
		\end{equation}
		\item With $U_{t-H+1:t}$ and $Y_{t-H+1:t}$, we then obtain the latent state $z_t$ using \eqref{eqn:z}. 
		\item For a prediction horizon of $T$ and with input $U_{t+1:t+T}$, let us recursively compute
		$$		\hat{Y}_{t+1:t+T}:= \{\hat{y}_{t+1},\hat{y}_{t+2},\cdots,\hat{y}_{t+T}\} $$
		from \eqref{eqn:z_next} and \eqref{eqn:haty}.
		\item The performance of the NSSM is evaluated through the following loss function
		\begin{equation}\notag
		\ell_{\text{SSM}} = \frac{1}{T}||Y_{t+1:t+T}-\hat{Y}_{t+1:t+T}||^2.
		\end{equation}
	\end{enumerate}
	
Obviously, $\ell_{\text{SSM}} $ is determined by both the dataset and the parameters of \eqref{eqn:SSM}. To make it clearer,  we explicitly define $\ell_{\text{SSM}}(\mD;\omega)$ as the loss function evaluated on the dataset $\mD$ and the parameter $\omega$.

\subsection{NSSM-enabled MPC }
Once trained, the NSSM can be used for predictive control tasks, such as within an MPC framework. To leverage this model to calculate the optimal input for the reference tracking, we rewrite \eqref{eqn:z_next} and \eqref{eqn:haty} in a compact form 
\begin{equation}\label{eqn:compact}
	s_{t+1} = A s_{t}+B u_{t+1},
\end{equation}
where \begin{equation}\notag
	s_{t+1}: = \begin{bmatrix}
		z_{t+1} \\ \hat{y}_{t+1}
	\end{bmatrix},\;
	A := \begin{bmatrix}
		A_z & 0\\
		C_zA_z & 0
	\end{bmatrix}, \; B := \begin{bmatrix}
		B_z\\C_zB_z
	\end{bmatrix}.
\end{equation}
With this linearized model, we propose to track the reference based on model predictive control (MPC). Notice that the choice of the controller is not limiting for the approach. Here we focus on MPC as it is a suitable approach for trajectory tracking under constraints. It solves the following problem:
	\begin{equation}\label{eqn:mpc}
		\begin{split}
			 \min _u \;&\tilde{y}_{t+N \mid t}^T \cdot P \cdot \tilde{y}_{t+N \mid t}\\&+\sum_{k=0}^{N-1} (\tilde{y}_{t+k \mid t}^T \cdot Q \cdot \tilde{y}_{t+k \mid t}+\Delta u_{t+k \mid t}^T \cdot R \cdot \Delta u_{t+k \mid t}) \\
				 \text { s.t. } \; &\eqref{eqn:SSM} \text{ and } u_{t+k \mid t} \in \mathcal{C}, \; k=0,\ldots, N-1,
		\end{split}
	\end{equation}
	where for $k=0,\ldots, N-1$, we define
	\begin{equation}\notag
	\begin{split}
		\tilde{y}_{t+k \mid t}&\triangleq \hat{y}_{t+k \mid t}-\bar{y}_{t+k},\\
		\Delta u_{t+k \mid t}&\triangleq u_{t+k+1 \mid t}-u_{t+k\mid t}.
	\end{split}
	\end{equation}
Moreover, $Q\geq 0$ and $R>0$ are design parameters kept fixed, and $P$ is the
	solution to the discrete-time algebraic Ricatti equation
\begin{equation}
\notag
P=A^T \left(P-P B \left(R+B^T P B\right)^{-1} B^T  P\right)  A+Q.
\end{equation}
	
Note that in \eqref{eqn:SSM}, the nonlinearity is just used to initialize an otherwise linear prediction model. Hence the MPC optimization problem is convex and  computationally tractable. The MPC finds the optimal controller that minimizes the cost function in \eqref{eqn:mpc}, thereby directing the predicted output $\hat{y}$ towards the reference signal $\bar{y}$.

\subsection{Meta-learning framework}

Collecting data is often expensive in many applications. Therefore, $\mathcal{D}_{\text{target}}$ is typically of limited size. Training with limited data can lead to poor prediction performance of NSSMs, which can adversely affect predictive control performance. To address this challenge, we propose a meta-learning framework. 
The framework is designed to acquire an NSSM, usually offline, that leverages data from multiple source systems that share similarities with the target system. 

For instance, the source system could be a numerical model or a digital twin of the target system, and is easily accessible. It could also be a system with modified physical parameters (e.g., different load, friction parameters, etc.). As another motivating example, suppose that we successfully control some manufacturing process governed by a nonlinear system, and have access to data from this system. However, the manufacturing requirements change, which corresponds to a change in the physical parameters in the system, and the controller performance needs to be adjusted. By using meta-learning, we propose a method that achieves this adjustment using very few actual measurements from the new system (the target system).

Assume that we have access to a dataset containing input and output trajectories generated by $N_s$ different source systems in the form of \eqref{eqn:sys}, where the $k$-th system is parameterized by some vector $\theta_k$ sampled from the distribution $\Theta$. It is important to know that $\theta_k$ need not to be known. We represent the source dataset as
\begin{equation}
	\mathcal{D}_{\text{source}}= \{(U(\theta^k, T^k), Y(\theta^k, T^k))\}_{k=1}^{N_s}:= \{\mathcal{D}^k\}_{k=1}^{N_s},
\end{equation}
where $T^k$ is the length of trajectory produced by the $k$-th system,  and $\theta_k\sim\Theta$ for each $k\in\{1,\cdots, N_s\}$.
Our objective is to learn an aggregated NSSM by leveraging the source data $\mathcal{D}_{\text{source}}$. This aggregated model should have the ability to be quickly adapted to other systems only using small datasets and without explicitly estimating $\theta$.

\section{Meta-Learning-Based MPC Design}\label{sec:alg}
We tailor the meta-learning algorithm iMAML \cite{rajeswaran2019meta} to address the problem of reference tracking on the target system. The framework consists of two phases: the meta-training phase learns the aggregated NSSM using $\mathcal{D}_{\text{source}}$, while the meta-inference phase adapts this model to the target system using $\mathcal{D}_{\text{target}}$.

\subsection{Bi-level optimization problem in meta-learning}
The phase of meta-training involves two loops: the outer-loop updates the parameters of the NSSM based on the performance across multiple source systems and the inner-loop adapts the model to individual source systems. This enables the meta-learning algorithm to learn from multiple systems and generalize well to new datasets.

\begin{figure}[!tb]
	\centering
	\includegraphics[width=0.9\linewidth]{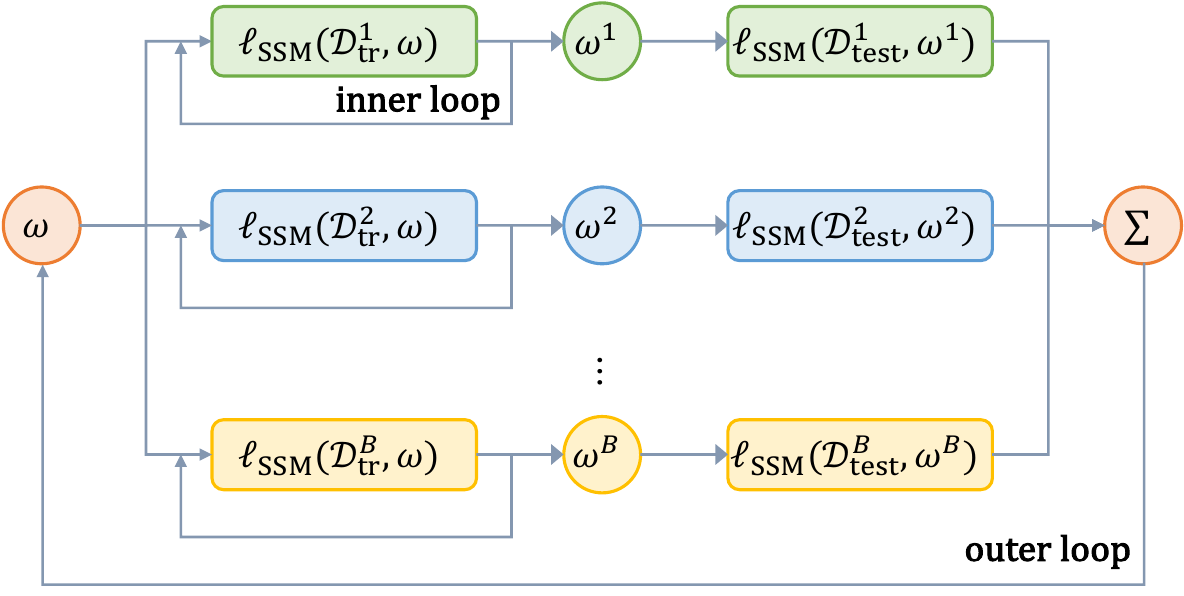}
	\caption{Information flow in the meta-training phase.}\label{fig:diag}
\end{figure}

Specifically, as shown in Fig.~\ref{fig:diag}, the outer loop provides an aggregated representation $\omega$ of the NSSM \eqref{eqn:SSM}. In each iteration of the outer loop, we sample a batch of trajectories $\{\mD^b\}_{k=1}^B$ from the source dataset $\mathcal{D}_{\text{source}}$. Then the inner-loop trains individual weights $\omega^b$ using the aggregated representation and the source data. After this, the outer loop updates the aggregated representation $\omega$ using these individual values. The process is repeated until convergence.

For each $\mD^b$, we partition it into a training set $\mD^b_{\text{tr}}$ and a testing set $\mD^b_{\text{test}}$. The training set is utilized to adapt the aggregated representation $\omega$ and produce the task-specific parameter $\omega^b$, tailored to the system \eqref{eqn:sys} parameterized by $\theta^b$. To be specific, within the context of iMAML, the inner-loop solves the following optimization problem:
\begin{equation}\label{eqn:inner_imaml}
	\omega^b(\omega):=\argmin_{\psi} \; \ell_{\mathrm{SSM}}(\mathcal{D}^b_{\text{tr}}; \psi)+\frac{\gamma}{2}||\psi-\omega||^2.
\end{equation}
Here, $\omega$ is the weight of the aggregated NSSM, and the regularization term $||\psi-\omega||^2$ encourages the finding of an optimal solution within an (ideally, small) neighborhood of $\omega$, and $\gamma>0$ controls the regularization strength. 
As we will shown later in Sections~\ref{sec:appro} and \ref{sec:maml}, this regularization enables us to express the outer-loop gradient term in closed-form by exploiting the implicit function theorem. 


The testing set $\mathcal{D}^b_{\text{test}}$ is employed in the outer loop to update the aggregated representation $\omega$. This is achieved by evaluating the prediction performance of the NSSM across multiple systems, as defined below
\begin{equation}\label{eqn:outer_prob}
\min_\omega \; \frac{1}{B}\sum_{b=1}^B \ell_{\mathrm{SSM}}(\mathcal{D}^b_{\text{test}}; \omega^b(\omega) ).
\end{equation}
As shown in \eqref{eqn:inner_imaml}, $\omega^b(\omega)$ is the optimal task-specific parameter adapted from $\omega$. Therefore, the objective of the outer loop is to learn a set of parameters $\omega$ that can produce good task-specific parameters after adaptation. For simplicity, we will denote $\omega^b(\omega)$ as $\omega^b$ in the rest of this paper.

\subsection{Solution to the bi-level optimization problem}\label{sec:appro}
We approach the solution of \eqref{eqn:outer_prob} using the gradient decent algorithm:
\begin{equation}\label{eqn:outer}
	\omega\gets \omega-\beta_{\text {out }} \frac{1}{B}\sum_{b=1}^B\nabla_\omega  \ell_{\mathrm{SSM}}(\mathcal{D}^b_{\text{test}} ; \omega^b),
\end{equation}
where $\beta_{\text {out }}$ denotes the learning rate for the outer loop. Considering \eqref{eqn:inner_imaml}, $\omega^b$ also depends on $\omega$. Therefore, by applying the chain rule, we expand the gradient term in \eqref{eqn:outer} as
\begin{equation}\label{eqn:meta_gradient}
	\nabla_\omega \ell_{\mathrm{SSM}}(\mathcal{D}^b_{\text{test}} ; \omega^b)= \frac{d \omega^b}{d \omega} \cdot P^b.
\end{equation}
Here,  $$P^b:=\nabla_\psi \ell_{\mathrm{SSM}}(\mathcal{D}^b_{\text{test}} ; \psi)|_{\psi=\omega^b},$$ 
which is the value of $\nabla_\psi\ell_{\mathrm{SSM}}(\mathcal{D}^b_{\text{test}}; \psi)$ evaluated at $\psi=\omega^b$. Since $P^b$ can be easily obtained via back propagation on the NSSM, we focus on calculating $\frac{d \omega^b}{d \omega}$.

Recall that $\omega^b$ is obtained by solving the inner optimization problem \eqref{eqn:inner_imaml}, typically through iterative algorithms like gradient descent. Thus, one approach to compute $\frac{d \omega^b}{d \omega}$ is propagating derivatives throughout the iterative process in the inner-loop. As this requires the full path of optimization to be stored in the memory, the approach becomes intractable if the number of iterative steps is large, or if the inner-loop procedure is non-differentiable. 
To address this challenge, the following lemma introduces an alternate method to calculate this gradient, which does not rely on the optimization path, but only on the optimal solution $\omega^b$.
\begin{lemma}[\!\!\cite{rajeswaran2019meta}]\label{lmm:implicit}
Define
\begin{equation}
Q^b:=I+\frac{1}{\gamma} \nabla_{\psi}^2 \ell_{\text{SSM}}(\mathcal{D}^b_{\text{tr}} ;\psi)|_{\psi=\omega^b},
\end{equation} 
where $ \nabla_{\psi}^2 \ell_{\text{SSM}}(\mathcal{D}^b_{\text{tr}} ;\psi)|_{\psi=\omega^b}$ is value of the Hessian  matrix $ \nabla_{\psi}^2 \ell_{\text{SSM}}(\mathcal{D}^b_{\text{tr}} ;\psi)$ evaluated at $\psi=\omega^b$.
If $Q^b$ is invertible, then 
\begin{equation}\label{eqn:Qb}
\frac{d \omega^b}{d \omega}=(Q^b)^{-1}.
\end{equation}
\end{lemma}

Combining \eqref{eqn:meta_gradient} and \eqref{eqn:Qb}, we obtain
\begin{equation}\label{eqn:gradient}
	\nabla_\omega \ell_{\mathrm{SSM}}(\mathcal{D}^b_{\text{test}} ; \omega^b) = (Q^b)^{-1} P^b.
\end{equation}
As $Q^b$ solely depends on $\omega^b$, Lemma~\ref{lmm:implicit} offers a solution for computing $\frac{d \omega^b}{d \omega}$ with reduced memory requirements. Nevertheless, obtaining $\omega^b$, the exact solution to the inner problem, involves solving \eqref{eqn:inner_imaml} until convergence, a task often impractical in practice. Therefore, in this paper, we resort to employing the gradient descent algorithm with a finite number of steps to obtain an approximate solution.

On the other hand, calculating the inverse of $Q^b$ can also be expensive, especially for large NSSMs. In order to tackle this issue, we notice that $(Q^b)^{-1}P^b$ is the solution to the following problem:
\begin{equation}\label{eqn:cg}
\min_{\phi} \; \phi^T Q^b \phi - \phi^T P^b.
\end{equation}
Therefore, in practice, we can also approximate $(Q^b)^{-1}P^b$ by solving \eqref{eqn:cg} with a finite number of iterations, e.g., update the solution by using a few gradient steps. By doing so, one avoids the need to explicitly compute the matrix inverse.

Finally, to ensure the compatibility with MPC, we update the source dataset online. After adaptation in each inner loop, we determine the optimal input $u^b_*$ by solving the MPC problem \eqref{eqn:mpc}. To balance exploration and exploitation, we let
\begin{equation}\label{eqn:ub}
	u^b = u^b_* + u^b_0,
\end{equation}
where $u^b_0 \sim \mathcal{N}(0, \Sigma^b)$. Injecting $u^b$ into \eqref{eqn:sys} (parameterized by $\theta^b$) yields an output $y^b$. We then update $\mathcal{D}_{\text{source}}$ by incorporating the most recent trajectory $(u^b, y^b)$. The complete meta-training algorithm is summarized in Algorithm~\ref{alg:meta_training}.

\begin{algorithm}
\small
	\caption{Meta-training phase}\label{alg:meta_training}
	\begin{algorithmic}
		\REQUIRE $\omega \leftarrow$ randomly initialize the parameters of the NSSM
		\REQUIRE $\mathcal{D}_{\text {source }} \leftarrow$ initial source dataset
		\REQUIRE Regularization strength $\gamma$ and learning rate $\beta_{\text{out}}$
		\WHILE[outer-loop]{not converge} 
		\STATE Sample batch $\{\mD^b\}_{b=1}^B$ from $\mathcal{D}_{\text {source }}$ 
		\FOR[inner-loop]{$b=1:B$}
		\STATE Partition $\mD^b$ into $\mathcal{D}^b_{\text{tr}}$ and $\mathcal{D}^b_{\text{test}}$
		\STATE $\omega^b \leftarrow$ solving \eqref{eqn:inner_imaml} by gradient descent
		\STATE $g^b \leftarrow$ solving \eqref{eqn:cg} by gradient descent
			\STATE $u^b\leftarrow$ the new input computed by \eqref{eqn:mpc} and \eqref{eqn:ub} with the NSSM parameterized by $\omega^b$
		\STATE $y^b\leftarrow$ simulating the system \eqref{eqn:sys} using $u^b$ with $\theta=\theta^b$  
		\STATE Update $\mathcal{D}_{\text{source}}$ by adding $(u^b,y^b)$
		\ENDFOR
		\STATE $\omega \leftarrow \omega-\beta_{\text {out }} \frac{1}{B}\sum_{b=1}^Bg^b$
		\ENDWHILE
		\ENSURE $\omega_{\infty} \leftarrow$ parameters obtained after meta-training phase
	\end{algorithmic}
\end{algorithm}

\subsection{Meta-inference}

After the meta-learning phase, we acquire the aggregated NSSM parameterized by $\omega_\infty$. The subsequent meta-inference phase adapts this aggregated model to accurately capture characteristics of the target system.

\begin{algorithm}
\small
	\caption{Meta-inference phase with tracking MPC}\label{alg:meta_inference}
	\begin{algorithmic}
		\REQUIRE $\omega_{\infty} \leftarrow$ parameters obtained after meta-training phase
		\REQUIRE $\mathcal{D}_{\text {target }} \leftarrow$ target dataset
		\REQUIRE $\gamma$ 
		\STATE $\omega^* \leftarrow$ solving \eqref{eqn:infer} by gradient descent \hfill \COMMENT{adaptation\;}
		\WHILE[reference tracking]{not done}  
			\STATE $u^*\leftarrow $ the optimal solution of \eqref{eqn:mpc} based on the NSSM parameterized by $\omega^*$
			\STATE Inject $u^*$ to the target system
			\ENDWHILE
	\end{algorithmic}
\end{algorithm}

Algorithm~\ref{alg:meta_inference} outlines the meta-inference process. We perform an adaptation step similar to \eqref{eqn:inner_imaml} to obtain the specific NSSM for the target system by using the limited dataset $\mathcal{D}_{\text{target}}$. This is expressed as
\begin{equation}\label{eqn:infer}
\omega^*:=\argmin_{\psi} \; \ell_{\mathrm{SSM}}(\mathcal{D}_{\text{target}}; \psi)+\frac{\gamma}{2}||\psi-\omega_\infty||^2.
\end{equation}
In practice, we approximate the optimal solution using only a few gradient steps. This corresponds to few ``trials'' to correctly adapt the parameters. With this NSSM tailored for the target system, i.e., the NSSM parameterized with $\omega^*$, we can achieve the reference tracking by using MPC \eqref{eqn:mpc}.


\subsection{Comparison with MAML}\label{sec:maml}
Algorithm~\ref{alg:meta_training} is developed based on the iMAML algorithm \cite{rajeswaran2019meta}. Alternatively, one could also consider here the MAML algorithm \cite{finn2017model}. Unlike \eqref{eqn:inner_imaml}, the inner loop of the MAML-based NSSM approximates the solution of
\begin{equation}\label{eqn:inner_maml}
	\omega^b:=\argmin_{\psi} \; \ell_{\mathrm{SSM}}(\mathcal{D}^b_{\text{tr}}; \psi).
\end{equation}
Namely, setting $\gamma$ in \eqref{eqn:inner_imaml} as $0$. In practice, the algorithm performs the adaptation through the gradient descent:
\begin{equation}\label{eqn:inner}
	\begin{split}
		\omega_{0}^b &= \omega,\\
		\omega_m^b&=\omega_{m-1}^b-\beta_{\text {in }} \nabla_{\omega_{m-1}^b} \ell_{\mathrm{SSM}}(\mathcal{D}^b_{\text{tr}} ; \omega_{m-1}^b), \;i = 1,\cdots, M,\\
		\omega^b &=\omega_M^b,
	\end{split}
\end{equation}
where $M$ is the number of inner-loop iterations and $\beta_{\text {in }}$ is the inner-loop learning rate. The outer-loop of MAML aims to solve the same problem as \eqref{eqn:outer_prob}. However, since a different cost function is optimized in \eqref{eqn:inner_maml}, $\frac{d \omega^b}{d \omega}$ cannot be directly calculated as in \eqref{eqn:Qb}. The exact calculation involves propagating derivatives along the optimization path, which is not memory-efficient. To tackle this, existing approaches typically approximate \eqref{eqn:outer} in the outer loop update as:

\begin{equation}\label{eqn:outer_maml}
	\omega \gets \omega - \beta_{\text {out }} \frac{1}{B}\sum_{b=1}^B\nabla_{\omega^b} \ell_{\mathrm{SSM}}(\mathcal{D}^b_{\text{test}} ; \omega^b),
\end{equation}
where it is assumed that
\begin{equation}\label{eqn:appro}
	\nabla_{\omega} \ell_{\mathrm{SSM}}(\mathcal{D}^b_{\text{test}} ; \omega^b) \approx \nabla_{\omega^b} \ell_{\mathrm{SSM}}(\mathcal{D}^b_{\text{test}} ; \omega^b).
\end{equation}
This approximation results in a tractable solution since $\nabla_{\omega^b} \ell_{\mathrm{SSM}}(\mathcal{D}^b_{\text{test}} ; \omega^b)$ can be readily obtained. However, \eqref{eqn:appro} may harm prediction performance of the NSSMs. In contrast, by leveraging the iMAML, our proposed algorithm can calculate the gradient without relying on the optimization path, see \eqref{eqn:gradient}. Therefore, there is no need to store the full optimization path or approximate the gradient as required in MAML. Consequently, the proposed approach reduces memory consumption and enhances control performance.

\section{Simulation}\label{sec:sim}
In this section, we illustrate the performance of our meta-learned MPC through some numerical examples.

\subsection{Van der Pol oscillators}
We first consider a family of van der Pol oscillators. The dynamics of each oscillator is given by
\begin{equation}
\begin{split}
\dot{x}_1&=x_2, \;
\dot{x}_2=\theta x_2\left(1-x_1^2\right)-x_1+u, \\
y&=x,
\end{split}
\end{equation}
where $\theta$ is the unknown damping ratio. By using meta-learning, we want the approach to work for various different damping ratios given the data of several of them. The source and target systems are generated by sampling $\theta$ from a Gaussian distribution $\theta\sim\mathcal{N}(0,1)$. 

We select the dimension of the latent state $z$ as $n_z = 5$. The encoder $f_{\text{enc}}$ is formed by a neural network comprising one input layer, one hidden layer, and one output layer, with each layer containing $128$ neurons and activated by rectified linear units (ReLUs). Notice that we can afford to use a shallow network here due to the explicit incorporation of the adaptation mechanism into the meta-training process. However, if we were required to learn a aggregated model for the entire distribution $\Theta$, we would need more depth and a larger-dimensional latent vector. The matrices in the SSM $A_z, B_z$ and $C_z$, are randomly initialized. The experiments are repeated for $10$ different random seeds.

\begin{figure}[!tb]
	\centering
	\includegraphics[width=0.7\linewidth]{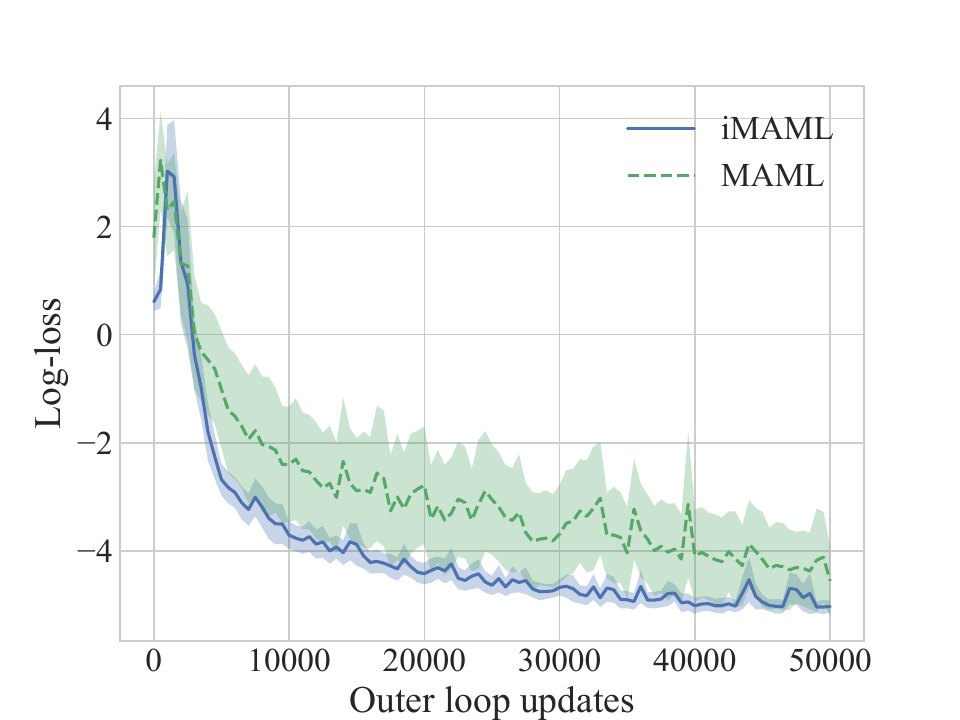}
	\caption{Comparison of prediction performance in the meta-training phase.}\label{fig:loss_compare}
\end{figure}

First, we evaluate the prediction performance of the NSSM during the meta-training phase. We compare the iMAML-based Algorithm~\ref{alg:meta_training}, with the MAML-based approach as outlined in Section~\ref{sec:maml}. The comparison is depicted in Fig.~\ref{fig:loss_compare}, where the $x$-axis refers to the number of outer loop updates, that is, \eqref{eqn:outer} and \eqref{eqn:outer_maml} respectively in the two algorithms. In each outer loop, we sample a batch of $B=16$ datasets from $\mathcal{D}_{\text{source}}$. Moreover, the performance is evaluated by taking the logarithm of $1/B\sum_{b=1}^B\ell_{\mathrm{SSM}}(\mathcal{D}^b_{\text{test}}; \omega^b)$, showing the prediction performance of the NSSMs. It is observed that Algorithm~\ref{alg:meta_training} outperforms the MAML-based approach throughout the training stage since it computes the derivative $\frac{d\omega^b}{d\omega}$ more accurately and thus requires less iterations for the same loss, as discussed in Section~\ref{sec:maml}. 

To evaluate the tracking performance on the target system, we compare against a few approaches, all of which have the same NSSM architecture but are initialized differently: 
\begin{enumerate}
	\item \textbf{iMAML}: The parameters of the NSSM are set with those obtained from the meta-training phase of the iMAML-based algorithm, that is, $\omega_{\infty}$ output by Algorithm~\ref{alg:meta_training}.
	\item \textbf{MAML}: 
	The parameters of the NSSM are set with those obtained from the meta-training phase of the MAML-based algorithm as outlined in Section~\ref{sec:maml}. 
	\item \textbf{Supervised learning}: Since there is no meta-training phase, the parameters are chosen randomly. This comparison with \textbf{Supervised learning} highlights whether meta-learning can improve our performance by leveraging the knowledge from source systems.
\end{enumerate}
Starting from these different initial parameters, we train the NSSMs using data from the target system $\mD_{\text{target}}$, which only contains $300$ data points. Specifically, for \textbf{iMAML}, this refers to run the adaptation step in Algorithm~\ref{alg:meta_inference}. On the other hand, \textbf{iMAML} and \textbf{supervised learning} perform the following update 
\begin{equation}
\begin{split}
\omega_m^*&=\omega_{m-1}^*-\beta_{\text {in }} \nabla_{\omega_{m-1}^*} \ell_{\mathrm{SSM}}\left(\mathcal{D}_{\text{target}} ; \omega_{m-1}^*\right),
\end{split}
\end{equation}
with $\omega_{0}^*$ being the parameters initialized for the NSSMs. 

\begin{figure}[!tb]
	\centering
	\includegraphics[width=0.7\linewidth]{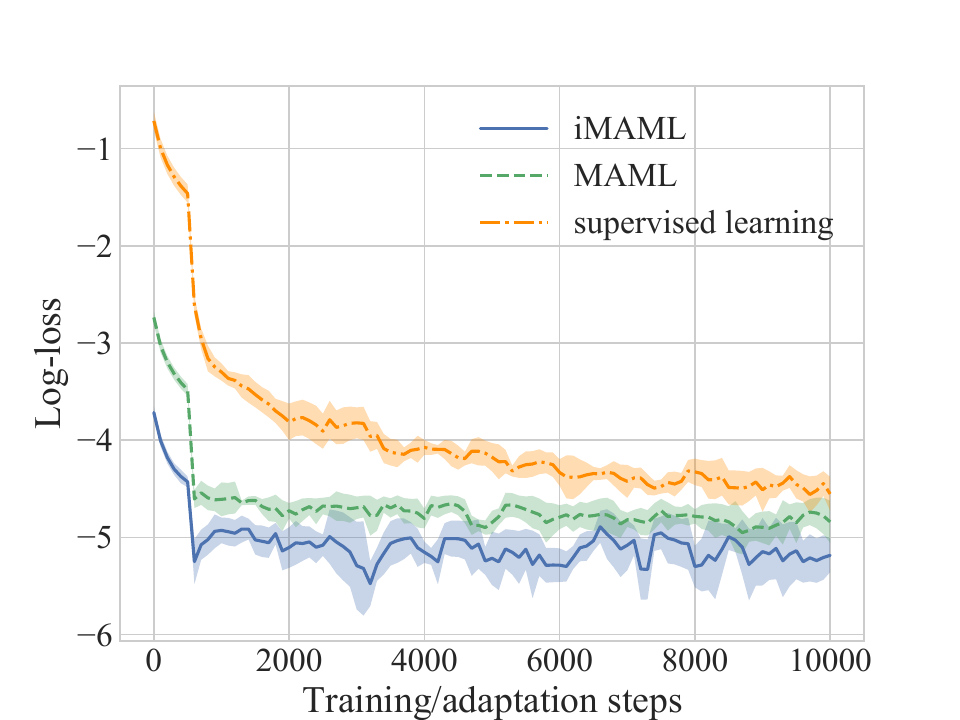}
	\caption{Comparison of prediction performance on the target system.}\label{fig:loss_compare_infer}
\end{figure}

It is evident from Fig.~\ref{fig:loss_compare_infer} that meta-learning expedites the training of NSSMs by using data from similar systems. The results reveal that \textbf{iMAML} tends to exhibit faster convergence and better tracking performance, especially in the initial stages of meta-inference. The enhancement is particularly striking when comparing iMAML and MAML tracking. This is because that \textbf{iMAML} exhibits superior performance with lower loss during the meta-training phase, as evidenced by Fig.~\ref{fig:loss_compare}. Consequently, it starts with better initial parameters, facilitating more effective adaptation to the target system.



\section{Conclusion}\label{sec:conclude}

This paper considers the problem of reference tracking in an unknown nonlinear system by using an NSSM. It introduces an iMAML-based MPC algorithm, which comprises two phases. In the meta-training phase, data from source systems is leveraged to pre-train the NSSM, while in the meta-inference phase, this model is quickly adapted to the target system using only a small amount of data. This approach addresses the issue of limited dataset availability for the target system. Numerical examples demonstrate the superior performance of the proposed algorithm compared to several baselines. As a future work, we are planning experiments on physical benchmark problems with a view towards transitioning to manufacturing processes.

\bibliographystyle{IEEEtran}\bibliography{reference} 
	
\end{document}